# Crystal Structure of BaFe$_2$Se$_3$ as a Function of Temperature and Pressure: Phase Transition Phenomena and High-Order Expansion of Landau Potential

Svitlyk V.[1,2*], Chernyshov D.[2], Pomjakushina E.[3], Krzton-Maziopa A.[3§], Conder K.[3], Pomjakushin V.[4], Pöttgen R.[1], Dmitriev V.[2]

1 Institut für Anorganische und Analytische Chemie, WWU Münster, D-48149, Münster, Germany
2 Swiss–Norwegian Beam Lines at ESRF, BP220, F-38043 Grenoble, France
3 Laboratory for Development and Methods, Paul Scherrer Institute, 5232 Villigen, Switzerland
[§] Present addr.: Faculty of Chemistry, Warsaw University of Technology, 00-664 Warsaw, Poland
4 Laboratory for Neutron Scattering, Paul Scherrer Institute, 5232 Villigen, Switzerland

Corresponding author e-mail: svitlyk@uni-muenster.de

**Abstract**

BaFe$_2$Se$_3$ (*Pnma*, CsAg$_2$I$_3$-type structure), recently assumed to show superconductivity at ~ 11 K, exhibits a pressure-dependent structural transition to the CsCu$_2$Cl$_3$-type structure (*Cmcm* space group) around 60 kbar, as evidenced from pressure-dependent synchrotron powder diffraction data. Temperature-dependent synchrotron powder diffraction data indicate an evolution of the room-temperature BaFe$_2$Se$_3$ structure towards a high symmetry CsCu$_2$Cl$_3$ form upon heating. Around 425 K BaFe$_2$Se$_3$ undergoes a reversible, first order isostructural transition, that is supported by the differential scanning calorimetry data. The temperature-dependent structural changes occur in two stages, as determined by the alignment of the FeSe$_4$ tetrahedra and corresponding adjustments of the positions of Ba atoms. On further heating, a second order phase transformation into the C*mcm* structure is observed at 660 K. A rather unusual combination of isostructural and second-order phase transformations is parameterized within phenomenological theory assuming high-order expansion of Landau potential. A generic phase diagram mapping observed structures is proposed on the basis of the parameterization.

1. Introduction

Detailed structural properties of novel superconducting materials are a key component for understanding of the phenomena. Polymorphism and phase transformations are of special interest as they bracket structural modifications where superconductivity may or may not exist. Enumeration of possible structural forms may be done with help of *ab-initio* calculations for many cases where stable structures correspond to various minima of the potential energy. Such an approach is, however, not an option for structures stabilized by a disorder or strongly anharmonic interactions. Such structures may exist at high temperatures and correspond to minima of the free energy that are difficult to deduce from an *ab-initio* structure calculated for 0 K. Phenomenological theory of phase transitions based solely on a symmetry basis helps to map those structures and relationships between them, but it has to be augmented by an experiment in order to constrain necessary polynomial expressions for the free energy and to orient phase diagrams in a space of physical parameters. Here we use a combination of diffraction experiments with phenomenological theory of phase transitions to uncover the unusual phase diagram of $BaFe_2Se_3$. For the title compound possible superconductivity at 11 K has been reported[1] thus manifesting a discovery of potentially new class of superconducting materials and calling for extended search of the possible polymorphs.

The crystal structure of $BaFe_2Se_3$ has been originally described in 1972[2]. It adopts the $CsAg_2I_3$-type structure, *Pnma* space group and is closely related to the $BaFe_2S_3$ (*Cmcm* space group; $CsCu_2Cl_3$ type[2]) structure. The structures of $BaFe_2S_3$ and $BaFe_2Se_3$ were considered as derivatives from the $SnNi_3$ structure [3] in term of distortions of close-packed hexagonal layers[4]. In this approach, the main structural building units are $BaS_{6/2}$ and $BaSe_{6/2}$ trigonal prisms. In $BaFe_2S_3$ and $BaFe_2Se_3$ the Ba cations are coordinated by eight chalcogen atoms, six of which form trigonal prisms. The other two atoms cap two rectangular faces of the trigonal prisms, with the corresponding Ba-(S/Se) distances equal in $BaFe_2S_3$ and unequal in distorted $BaFe_2Se_3$. As a result of such a deformation, the $BaFe_2Se_3$ structure loses *C*-centering and adopts the *Pnma* symmetry.

Since the discovery, physical properties of $BaFe_2Se_3$ have not been explored. Recently we have reported the synthesis, structural and magnetic properties of newly grown single crystals of the title compound[1]. They were found to undergo an antiferromagnetic transition below 240 K and exhibited a diamagnetic response around 11 K. Although the binary FeSe phase was observed as an impurity (< 0.8 mass %), bulk FeSe shows superconductivity at ~8.5 K [5-8] that is below the observed diamagnetic response. Following independent work indicated 255 K as the onset temperature for antiferromagnetic ordering with a diamagnetic signal around 10 K, the sample was reported to contain the FeSe impurity as well[9]. Other studies on $BaFe_2Se_3$ indicated 256 K as a Néel temperature and reported an absence of superconductive response down to 1.8 K[10]. Interestingly, the iron deficient semiconducting

BaFe$_{1.79(2)}$Se$_3$ phase did not exhibit long-range magnetic order in the 1.8 – 400 K temperature range[11]. Instead a spin glass-like behavior was observed at about 50 K. Controversy in the literature data could originate from the fact that physical properties of BaFe$_2$Se$_3$ strongly depend on the exact composition of the studied samples.

Here we further characterize BaFe$_2$Se$_3$ with the help of single crystal and powder diffraction of synchrotron radiation as a function of temperature and pressure. At variance with a previously proposed description in terms of BaSe$_{6/2}$ trigonal prisms, we parameterize the observed structural evolution as correlated rotations of relatively rigid double chains of shared FeSe$_4$ tetrahedra accompanied by a displacement of Ba cations. Such a description directly follows from our structural data and is further supported by a symmetry-based analysis in terms of deformation modes. We show that the transition between the two phases of the *Cmcm* and *Pnma* symmetries is induced by applying ~7 GPa pressure at room temperature. Under constant ambient pressure similar transition is observed on heating at 660 K. We also found a first-order isostructural phase transition occurring around 425 K. Rather unusual combination of phase transformations calls for high order terms in Landau expansion of free energy; we propose a corresponding phenomenological model and validate it with experimental observations.

## 2. Experimental

Single crystals of the BaFe$_2$Se$_3$ sample studied in this work are the same as in Ref. [1]. The crystals were grown from a melt of high purity elements (at least 99.99%, Alfa) using the Bridgman method. Detailed description of the synthetic procedure can be found in our original work [1].

### 2.1. Differential scanning calorimetry

Differential scanning calorimetry (DSC) experiments were performed with a Netzsch DSC 204F1 system. Measurements were performed on heating and cooling with a rate of 10 K/min using a 20 mg sample encapsulated in a standard Al crucible. An argon stream was used during the experiment as a protecting gas. The experimental DSC curve is presented on Fig. 1.

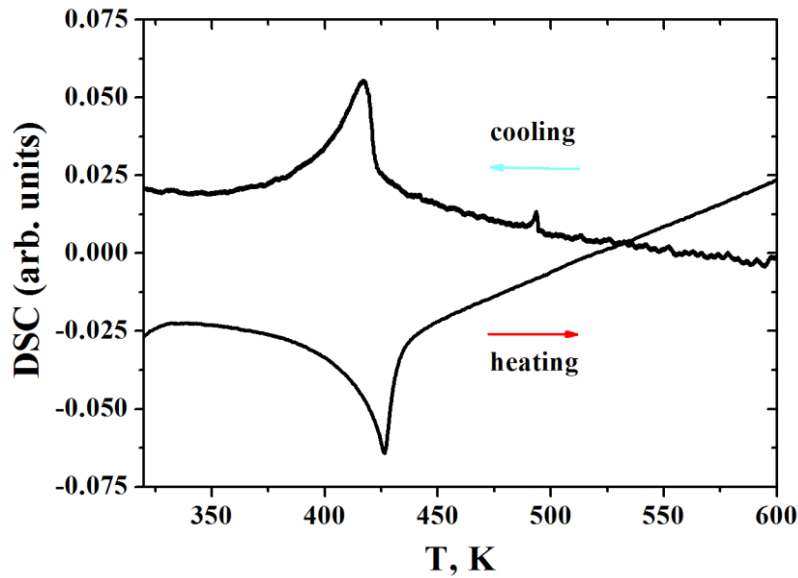

Figure 1. Differential scanning calorimetry curve showing a first-order transition with a hysteresis around 425 K.

### 2.2. Temperature- and Pressure-Dependent X-ray Diffraction

High-temperature single crystal data were collected with the MAR345 image plate detector (IP) at the Swiss-Norwegian Beam Lines, BM01A station at the ESRF (Grenoble, France) at 480 K using the wavelength of 0.69736 Å. For the collection at 480 K the detector was moved to a distance of 250 mm from the sample in order to obtain relatively high resolution low angle reciprocal space maps. The resulting crystallographic resolution for the 480 K setup is 1.17 Å. A Cryostream 700+ $N_2$ blower was used to control the temperature. Data were processed using the CrysAlis software[12]. The reciprocal layers of 480 K and together with the original data collected at 150 K[1] (Fig. 2) indicate close similarity in spite of the fact that they were collected above and below the transition observed with DSC. The difference in the signal to noise ratios observed on the experimental data sets (Fig. 2) originates from the type of the used detector. The presented 150 K data were collected with a CCD detector. Contrary to IP-based systems, CCD detectors offer fast readout time but feature intrinsic electronic noise, therefore they accumulate an additional detector-related background during the data collection. IP detectors allow to collect data with high signal to noise ratio, thus allowing to observe weaker features of reciprocal space in greater details; the mosaicity in the $BaFe_2Se_3$ sample is much better observed at the 480 K data set for the same reason.

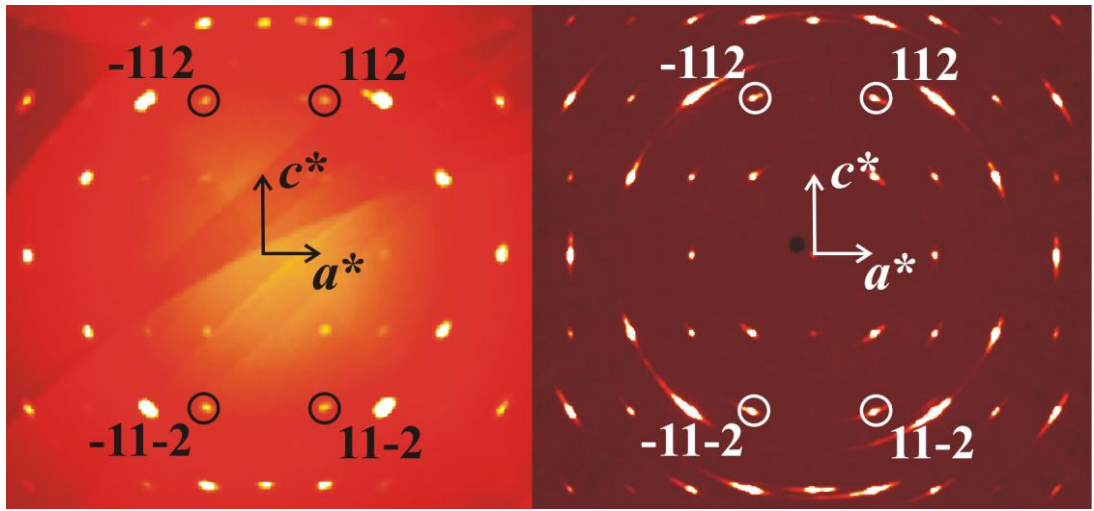

Figure 2. Reciprocal $h1l$ layers for $BaFe_2Se_3$ from the single crystal data collected at 150 (left) and 480 K (right).

For the temperature-dependent powder diffraction a small amount of the $BaFe_2Se_3$ single crystals was finely ground and placed into a glass capillary. Initially data were collected with the MAR345 detector in a transmission mode using the same wavelength of 0.69736 Å. The temperature was changed in a ramp mode with parallel collection of powder patterns from 80 to 500 K on heating and from 500 down to 360 K on cooling to check for a possible hysteresis using the Cryostream 700+ $N_2$ blower. Higher temperatures of up to 960 K were reached with an inhouse developed heat blower. The high-temperature data were collected in a ramp mode using a Dectris Pilatus2M detector. Experimental data around 660 K are shown on Fig. 3.

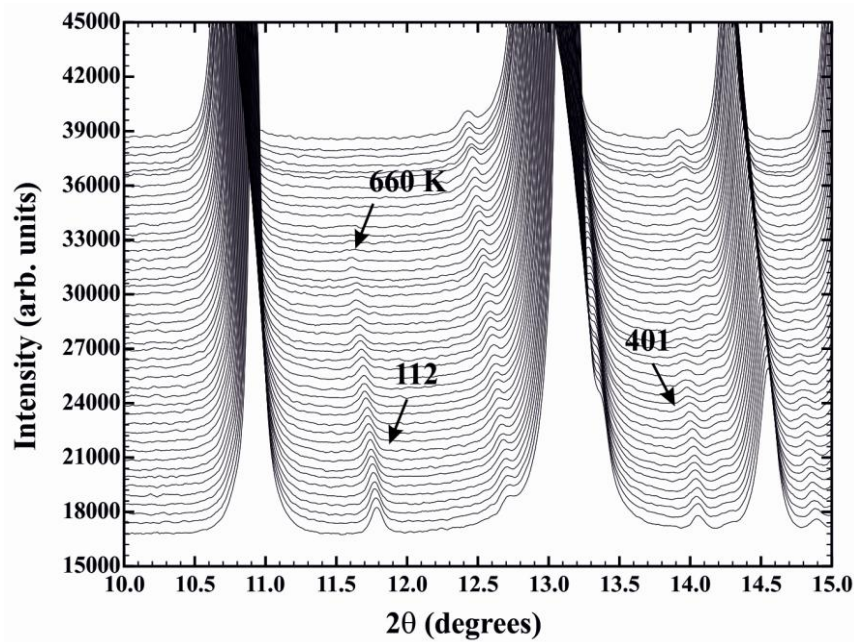

Figure 3. Temperature-dependent powder diffraction scans for $BaFe_2Se_3$.

For the pressure-dependent powder diffraction, the BaFe$_2$Se$_3$ single crystals were finely ground and loaded into the high pressure diamond anvil cells. An ethanol-methanol mixture in a 1:4 ratio was used as a pressure transmitting medium. For the powder diffraction data collection pressure was changed from ambient to 160 kbar with a typical step of 15 kbar. The sample was fixed by the stainless steel gasket with a hole of 0.3 mm in diameter. The effective pressure was calculated from the shift of the fluorescence signal from the ruby crystals also loaded in the cell. The experimental data set is presented on Fig. 4.

Both temperature- and pressure-dependent powder diffraction data were integrated and processed with the FIT2D software[13], standard deviations were calculated with a home-made software. Structural parameters were refined with the FullProf software[14].

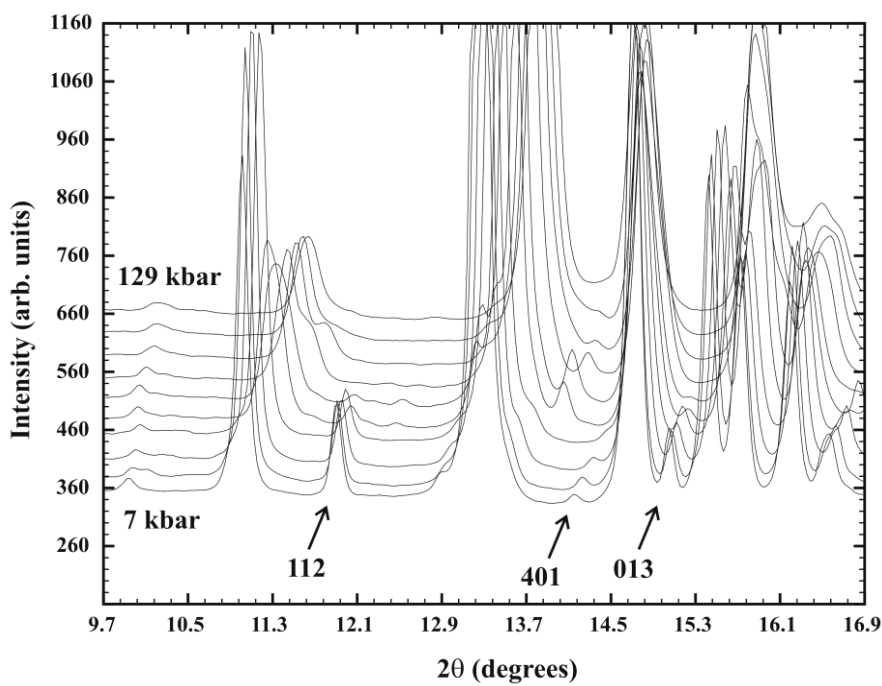

Figure 4. Pressure-dependent powder diffraction scans for BaFe$_2$Se$_3$.

### 3. Results

We start from the high-pressure experiment in DAC which indicates a structural phase transition at ~ 60 kbar (Fig. 4). The data agree with the expected *Pnma* → *Cmcm* symmetry change, as can be noticed in particular from the 112, 401 and 013 reflections that fully disappear between 51.5 and 70.5 kbar (Fig. 5); those reflections are allowed in *Pnma* but forbidden in *Bbmm* space group (non-standard setting) if we preserve the *Pnma* unit-cell setting, or in *Cmcm* space group (standard setting) with axes relating to *Pnma* with a help of the [010 001 100] matrix (non-standard setting *Pbnm*). We denote this high-symmetry polymorph, isostructural to $BaFe_2S_3$, as γ-$BaFe_2Se_3$. The small number of data points and the limited quality of the high pressure powder data do not allow to make firm conclusions on the order of the transition, however, close to linear dependence of the intensity of discriminating reflections on pressure is indicative of a second order phase transition (Fig. 5).

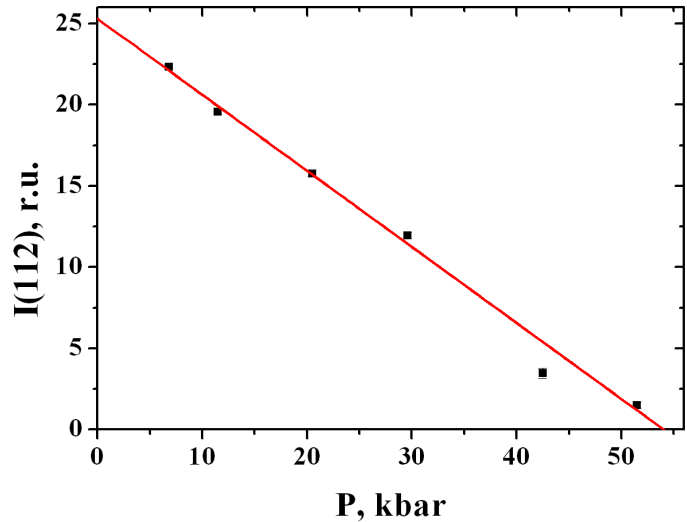

Figure 5. Least-square intensity fit of the 112 peak series from the pressure-dependent powder diffraction data.

Extrapolation of the linear fit (Fig. 5, solid red line) suggests 54 kbar as a transition pressure. Volume vs. pressure dependences for α- and γ-$BaFe_2Se_3$ were fitted with the Murnaghan equation of state [Eq. (1), $V_0$ is the volume at zero pressure, $B_0$ is the bulk modulus and $B_0'$ is the first pressure derivative of the bulk modulus] and are shown on Fig. 6. Fitted parameters are presented in Table 1. Standard deviations for the parameters of γ-$BaFe_2Se_3$ are rather high due to a limited number of available data points.

$$V(P) = V_0(1 + B_0'\frac{P}{B_0})^{-\frac{1}{B_0'}} \quad (1)$$

Table 1. Experimental coefficients of the Murnaghan equation of state for α- and γ-$BaFe_2Se_3$ from the pressure-dependent powder data

| sample | $V_0$, Å$^3$ | $B_0$, kbar | $B_0'$ |
|---|---|---|---|
| α-$BaFe_2Se_3$ | 589.3(8) | 444(23) | -0.7(9) |
| γ-$BaFe_2Se_3$ | 545(4) | 951(131) | 0(1) |

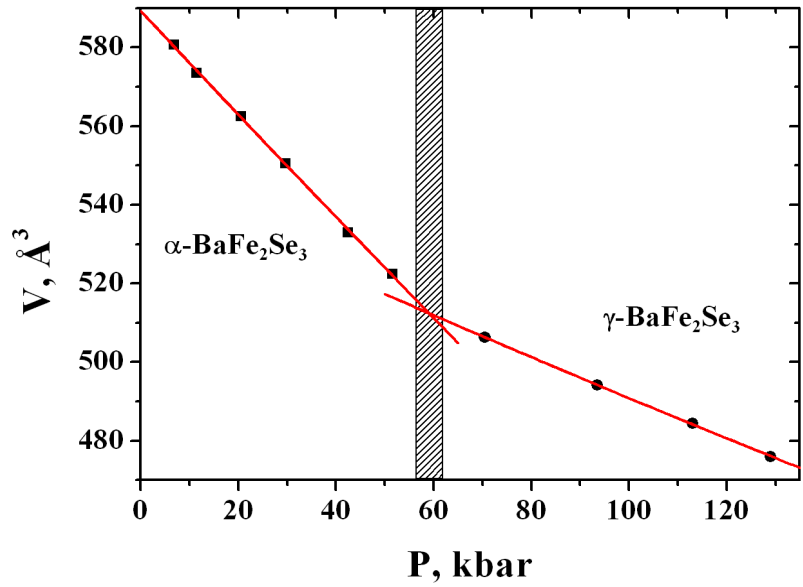

Figure 6. Volume vs. pressure dependences for the α- and γ-BaFe$_2$Se$_3$ phases fitted with the Murnaghan equation of state (1). The dashed bar indicates the expected position for the phase boundary.

Structural data for the *Cmcm* polymorph as they follow from the Rietveld fit (Fig. 7, *Cmcm* space group, $a$ = 8.7800(12), $b$ = 10.885(3), $c$ = 5.2981(6) Å, $R_p$ = 3.65, $R_{wp}$ = 4.93, $R_B$ = 5.76; impurity peaks correspond to the binary FeSe phase, < 0.8 mass %) of the 70 kbar data set are shown in Table 2. The refined structure has been compared to the low symmetry phase with help of ISODISTORT software [15]. Such an analysis allows to describe a low symmetry structure as a distorted version of the parent high symmetry analogue. The distortions are split into a finite set of contributions (symmetrical modes) induced by irreducible representations of the parent structure space group. The full comparison of two structures in terms of symmetrical modes together with their definitions are presented in the Supplementary material, Tables 1 and 2. The modes responsible for the transition comprise rotations of the Fe(Se/S)$_4$ tetrahedra together with displacements of Ba cations. In γ-BaFe$_2$Se$_3$ (Fig. 8, on the left, projection along the *c* axis is shown) the FeSe$_4$ tetrahedra are aligned in the horizontal plane. The α-BaFe$_2$Se$_3$ ambient pressure structure (Fig. 8, on the right, projection along *b* axis is shown) is obtained through a tilting of the tetrahedra along the γ-BaFe$_2$Se$_3$ *c* axis (BaFe$_2$Se$_3$ *b* axis); the tetrahedra located in the unit cell are rotated counterclockwise while the tetrahedra on the cell edges are rotated clockwise (Fig. 8).

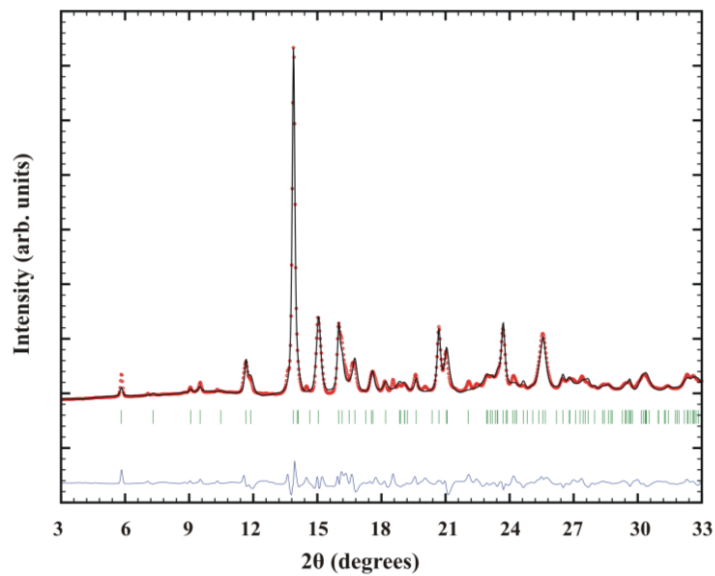

Figure 7. Rietveld refinement of the BaFe$_2$Se$_3$ powder synchrotron X-ray diffraction data at 70 kbar. Red dotted line corresponds to the experimental profile, solid black line corresponds to the calculated profile, vertical bars represent the Bragg positions and the bottom solid line is the difference profile.

Table 2. Structural data for the *Cmcm* polymorph of BaFe$_2$Se$_3$ at 70 kbar

| Atom | site | x | y | z | $B_{iso}$, Å$^2$ |
|---|---|---|---|---|---|
| Ba | 4c | 1/2 | 0.1741(8) | 1/4 | 0.11(13) |
| Se(1) | 4c | 1/2 | 0.637(1) | 1/4 | 0.16(14) |
| Se(2) | 8g | 0.2029(5) | 0.3785(9) | 1/4 | 0.14(14) |
| Fe | 8e | 0.349(1) | 1/2 | 0 | 2.33(19) |

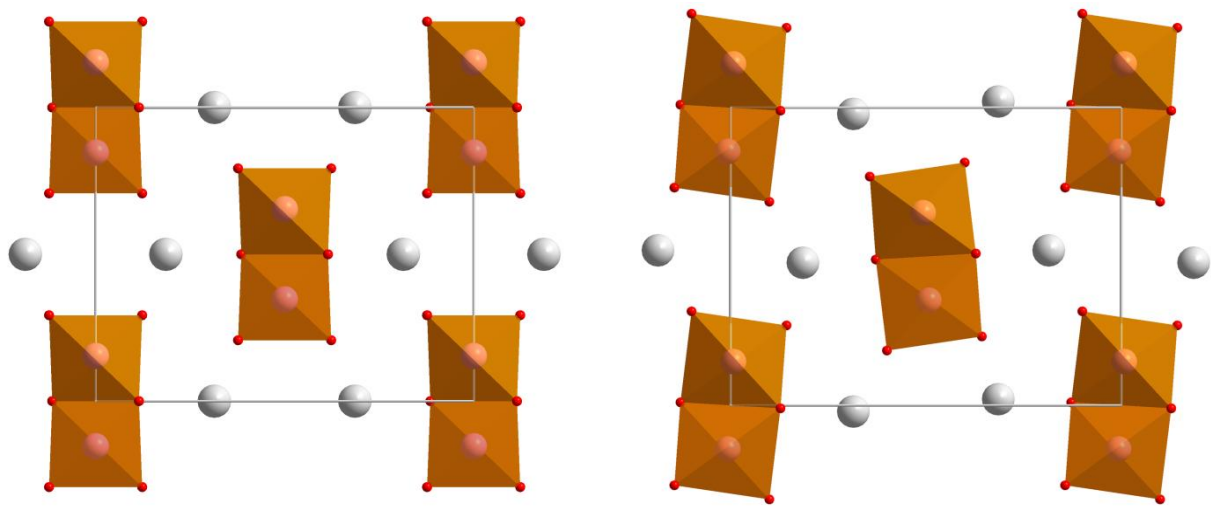

Figure 8. Transformation from the γ-BaFe$_2$Se$_3$ (left) to α-BaFe$_2$Se$_3$ (right) structure through a rotation of the FeSe$_4$ tetrahedra.

The rotations of the FeSe$_4$ tetrahedra during the transition induce changes in the environment of Ba atoms as well. These changes mirror the differences in the coordination of the Ba atoms in BaFe$_2$S$_3$ and BaFe$_2$Se$_3$, described in the **Introduction** part of the paper. The Ba cations are coordinated by eight Se atoms, six of which form trigonal prisms. The other two atoms which cap two rectangular faces of the trigonal prisms have the corresponding Ba-Se distances equal in γ-BaFe$_2$Se$_3$ and unequal in distorted α-BaFe$_2$Se$_3$.

The structural changes between the ambient- and high-pressure modifications of BaFe$_2$Se$_3$ are strictly related by a group-subgroup scheme which is presented in the Bärnighausen [16-19] formalism in Fig. 9. The *klassengleiche* symmetry reduction of index 2 (k2) from *Cmcm* to *Pbnm* (non-standard setting of *Pnma*) leads to emergence of superstructure reflections for the ambient-pressure phase. The latter is isotypic with RbCu$_2$I$_3$ [20, 21]

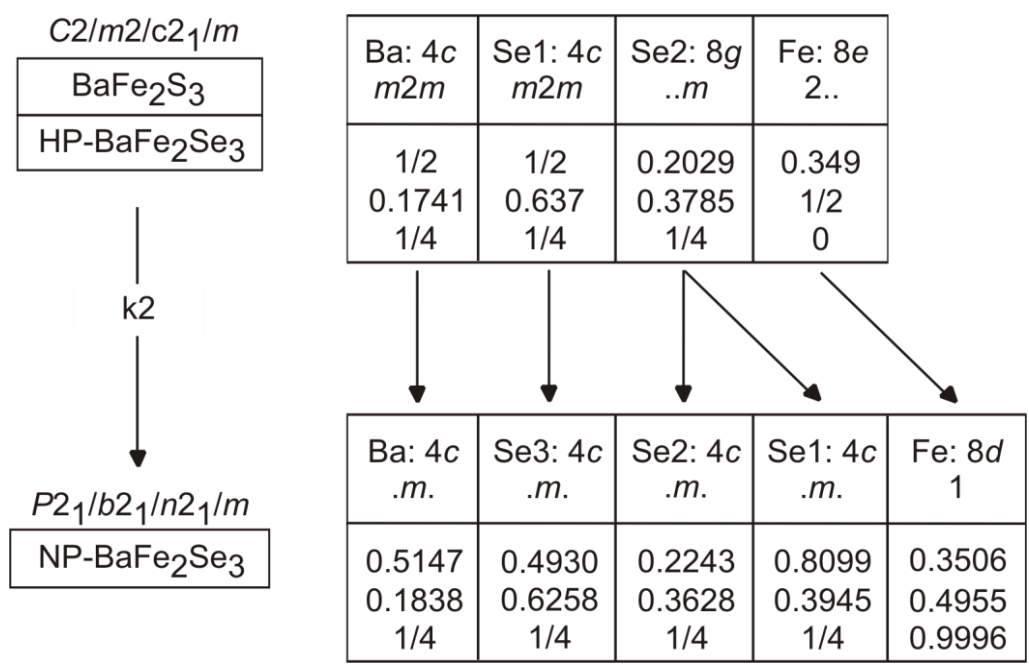

Figure 9. Group-subgroup scheme in the Bärnighausen formalism for the ambient- and high-pressure modifications of BaFe$_2$Se$_3$. The index for the *klassengleiche* symmetry reduction (k) and the evolution of the atomic parameters is given.

The symmetry reduction leads to a reduction of the site symmetry for the Ba, Se1, and Fe sites and a splitting of the 8*g* Se2 site into two fourfold 4*c* sites. The resulting selenium sites Se2 and Se1 have distinctly different *x* and *y* parameters as compared to the high symmetry phase. This allows the rotation of the tetrahedral units. The iron atoms at the centres of the tetrahedra remain almost at the ideal positions.

This kind of phase transition seems to be very sensitive to changes in the atom size. Applying pressure on the selenide one obtains the structure of the sulphide. Such displacive phase transitions can also be expected for the alkali metal halocuprates $ACu_2X_3$ ($A$ = Rb, Cs; $X$ = Br, I). Most of these halides adopt the *Cmcm* structure at room temperature and might show the *Pnma* phase upon cooling.

The characteristic peak on the experimental DSC data (Fig. 1) indicates a first-order phase transition at 426 K on heating and at 415 K on cooling. The observed temperature hysteresis of 11 K also agrees with a first-order character of the transition. Since the 150 K and room temperature structures have been reported before, we start from the single crystal diffraction experiment at 480 K in order to determine the features of the $BaFe_2Se_3$ structure above the transition. Visual inspection of the raw single crystal diffraction data does not reveal any apparent differences between the 150 and 480 K datasets (Fig. 2, $h1l$ layers are shown as an example). The quasi 1D structure along the $b*$ direction results in a mosaicity in the $a*c*$ planes, which is more clearly observed at 480 K as a result of an improved signal to noise ratio of the IP detector compared to the CCD detector and a high resolution experimental setup, as was mentioned in the **Experimental** section of the manuscript.

Transformation from the low-temperature α-$BaFe_2Se_3$ structure (*Pmna*) to a more symmetrical γ-phase (*Cmcm*) on heating could be expected similar to the pressure induced transformation. However, the experimental 480 K pattern indicates the persistence of a primitive unit cell. For instance, presence of the 112 reflections (Fig. 2), which would violate possible *C*-centering (or *B*-centering for the original *Pnma* unitcell setting), indicates that the 480 K structure is not transformed into the $BaFe_2S_3$ *Cmcm* structure. Therefore the first order phase transition detected by the DSC is not related to the expected polymorphism and detailed structural analysis is required to uncover its nature.

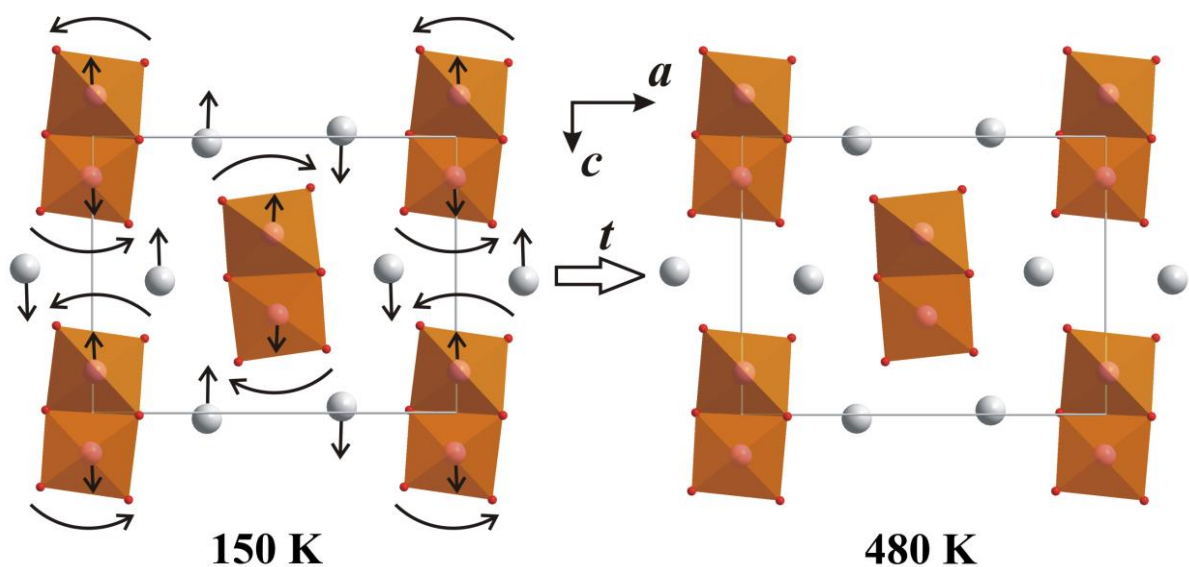

Figure 10. Evolution of the BaFe$_2$Se$_3$ *Pnma* structure with temperature.

Structural refinement of the single crystal data shows that the *Pnma* structure does evolve towards the *Cmcm* symmetry (Fig. 10), albeit the undistorted *Cmcm* structure is evidently not reached at 480 K. Comparison of bond lengths and bond angles for the 150 K and 480 K experiments indicates that the originally proposed key structural elements - BaSe$_{6/2}$ prisms - show strong deformation with temperature. At variance, FeSe double chains behave as nearly rigid blocks. Thus both the shape and size of the Ba coordination polyhedra change from 150 to 480 K with an increase of the Ba-Se distances averaged over 3 NN by ~ 0.064 Å while the Fe-Se distances averaged over 4 NN show only a 0.01 Å change. Albeit crystal's quality precludes data collection suitable for proper analysis of charge density and related bonding, our observations agree with stronger directional bonding within double chains, and weaker, presumably ionic bonding between chains and the barium atoms.

Similar conclusions are drawn from the analysis of symmetry-allowed deformations done with the ISODISTORT software[15]. The analysis is based on the comparison for two structures determined at 150 and 480 K and parameterization of the observed differences in terms of distortion modes induced by irreducible representations of the space-group symmetry (Supplementary Material, Table 3; Displacive mode definitions are presented in Table 4), here we note that the core structural difference between two data sets is the rotation of the FeSe$_4$ tetrahedra within the *ac* plane. The Ba atoms adjust their positions following their eight-fold coordination by the Se atoms; the shifts take place mainly along the crystallographic *c* direction with a minor component along the *a* direction. The Fe atoms serve as rotation centers for the FeSe$_4$ tetrahedra and, consequently, undergo only slight displacements. All atomic movements take place solely in the *ac* plane (Fig. 10). Note that all the distortions do not change the symmetry, in agreement of with the experiment. Results of structural refinement and crystal data for the BaFe$_2$Se$_3$ single crystal collected at 480 K are presented in Tables 3 and 4.

Table 3. Single crystal data and structure refinement for BaFe$_2$Se$_3$ ccollected at 480 K

| Empirical formula | BaFe$_2$Se$_3$ |
|---|---|
| Wavelength, Å | 0.69736 |
| Space group | *Pnma* |
| Lattice parameters *a*/*b*/*c*, Å | 11.9700(3)/5.4976(2)/9.2435(2) |
| Volume, Å$^3$ | 608.28(3) |
| Z | 6 |
| Density (calculated), g/cm$^3$ | 7.96 |
| 2$\theta$ range for data collection, ° | 6.68 - 34.56 |
| Index ranges, *h*; *k*; *l* | ±10; ±4; ±7 |
| Reflections collected | 2443 |
| Independent reflections | 208 [$R_{int}$ = 0.0369] |
| Completeness to max 2$\theta$, % | 92.0 |
| Data/parameters | 208/35 |
| Goodness-of-fit on $F^2$ | 0.783 |
| Final R indices [$I > 2\sigma(I)$] | $R_1$ = 0.0310, w$R_2$ = 0.1042 |
| R indices (all data) | $R_1$ = 0.0310, w$R_2$ = 0.1042 |
| Extinction coefficient | 0.003(3) |
| Largest diff. peak/hole, e/Å$^3$ | 0.78/-0.73 |

Table 4. Atomic coordinates and equivalent isotropic displacement parameters ($U_{eq}$) for the BaFe$_2$Se$_3$ single crystal collected at 480 K

| Atom | Site | *x* | *y* | *z* | $U_{eq}$, Å$^2$ |
|---|---|---|---|---|---|
| Ba | 4*c* | 0.1838(1) | 1/4 | 0.5147(1) | 0.037(1) |
| Se(1) | 4*c* | 0.3945(2) | 1/4 | 0.8099(2) | 0.034(1) |
| Se(2) | 4*c* | 0.3628(2) | 1/4 | 0.2243(2) | 0.031(1) |
| Se(3) | 4*c* | 0.6258(1) | 1/4 | 0.4930(1) | 0.020(1) |
| Fe | 8*d* | 0.4955(1) | 0.9996(2) | 0.3506(2) | 0.018(1) |

Therefore the transition observed with DSC around 425 K (Fig. 1) is isostructural in the sense that two phases are of the same space group symmetry and corresponding atoms occupy the same Wyckoff positions and thus must be of a displacive type.

Probing structural behavior at the transition and throughout whole temperature range by means of single crystal analysis is time- and resource-consuming. Therefore synchrotron X-ray powder diffraction using a 2D image plate data recording was employed to probe the structure in fine steps throughout a wide temperature range.

Series of experimental temperature-dependent powder patterns indicate that the observed 112, 401, 013 reflections that would violate *C*-centering of the possible high-temperature $BaFe_2S_3$ *Cmcm* structure (*Bbmm* space group in the *Pnma* unit-cell setting) do not disappear up to the temperature of 500 K, in agreement with the single crystal data. Careful inspection reveals nearly linear decrease of their intensities with heating (intensities of 112 reflections are shown on Fig. 13, black datapoints).

Powder diffraction data were fitted with the *Pnma* model in the full temperature range from 80 to 500 K. Anomalies are present in the behavior of all structural parameters, with the most prominent for the *a* cell parameter (Fig. 11, left) and the *z* coordinates of the Se atoms (Fig. 11, right, the Se3 *z* coordinate is shown). Changes in the unit-cell volume with temperature are presented on Fig. 12.

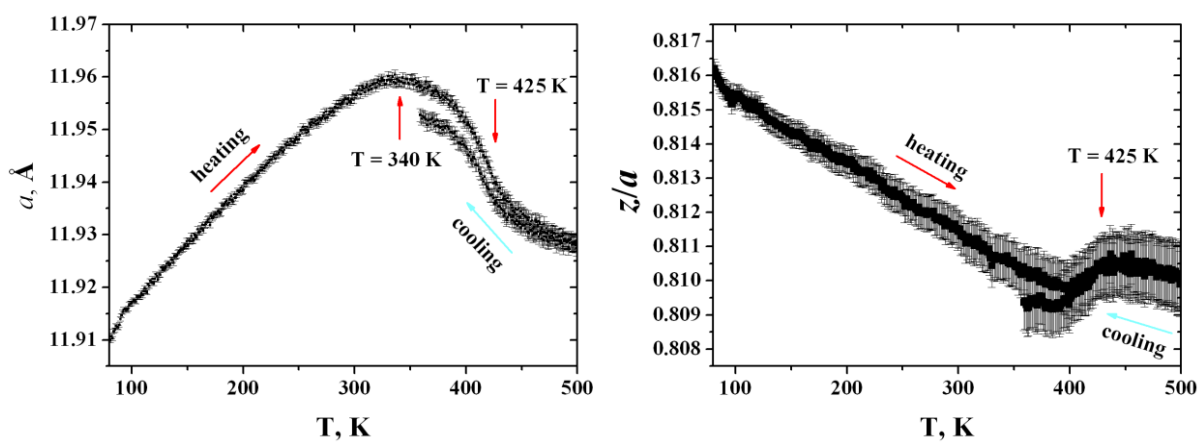

Figure 11. Dependence of the *a* unit-cell parameter (left) and Se3 *z* coordinate (right) from temperature.

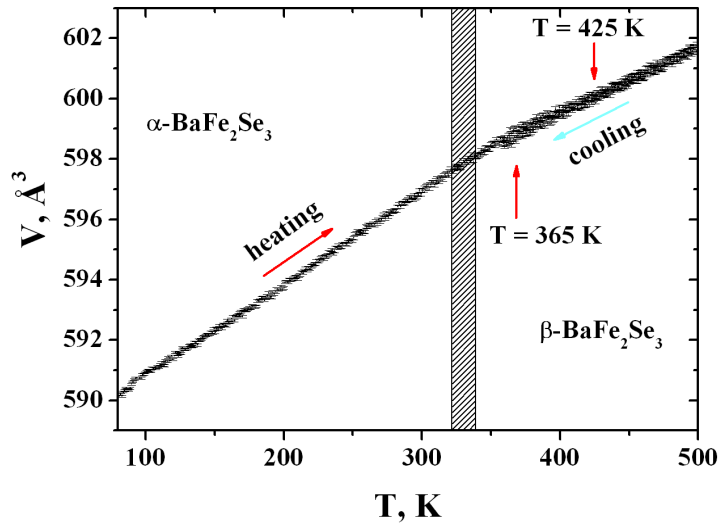

Figure 12. Temperature dependence of the unit cell volume. The dashed bar indicates the phase boundary.

As seen from the temperature-dependent behavior of the *a* unit-cell parameter (Fig. 11, left), structural changes commence at the temperature of 340 K. Around 425 K the *a* unit-cell parameter undergoes rapid decrease and the Se *z* coordinates experience sudden jumps (Fig. 11, right, Se3 *z* coordinate is shown). The structural changes at 425 K are abrupt enough to exhibit first-order response on the DSC curve (Fig. 1). The only anomaly at the temperature-dependent behavior of the unit-cell volume, if any, may be seen around 365 K. Hysteresis in the structural parameters further supports first-order nature of the transition.

The temperature evolution of the $BaFe_2Se_3$ structure can therefore be tentatively divided into two regions: from 80 to about 340 K, and from 340 to 500 K. In both temperature regions the transition is manifested as a rotation of the $FeSe_4$ tetrahedra with corresponding adjustments of the Ba positions, as concluded from the simulations using the ISODISTORT tool[15]. While in the first temperature region both rotations of the $FeSe_4$ units and movements of Ba atoms are near equally involved in the transition, the structural changes in the 340 - 500 K region are dominated by a displacement of the Ba atoms. We will denote the low-temperature (below 340 K) $BaFe_2Se_3$ phase as α and the high-temperature (from 340 to at least 500 K) one as β. The two phases have the same space group symmetry but feature a small difference in coordination and chemical bonding, a consequence of change in the lattice parameters.

Upon further increase of the temperature the *Pnma-Cmca* transition is observed around 660 K (Fig. 3). Dependence of the intensity of the 112 reflection with temperature is shown on Fig. 13. Black datapoints correspond to the data obtained with the cryostream blower, red datapoints correspond to the data obtained with the heat blower. Refined structural parameters of the $BaFe_2Fe_3$ $Cmcm$ modification at 660 K are presented in Table 5 ($Cmcm$ space group, $a = 9.2300(3)$, $b = 11.9088(4)$, $c = 5.5022(1)$ Å, $R_p = 7.62$, $R_{wp} = 8.07$, $R_B = 3.13$)

Table 5. Structural data for the $Cmcm$ polymorph of $BaFe_2Se_3$ at 660 K

| Atom | site | x | y | z | $B_{iso}$ |
|---|---|---|---|---|---|
| Ba | 4c | 1/2 | 0.1818(2) | 1/4 | 2.41(4) |
| Se(1) | 4c | 1/2 | 0.6257(3) | 1/4 | 1.54(4) |
| Se(2) | 8g | 0.2067(2) | 0.3780(2) | 1/4 | 1.54(4) |
| Fe | 8e | 0.3545(3) | 1/2 | 0 | 1.15(4) |

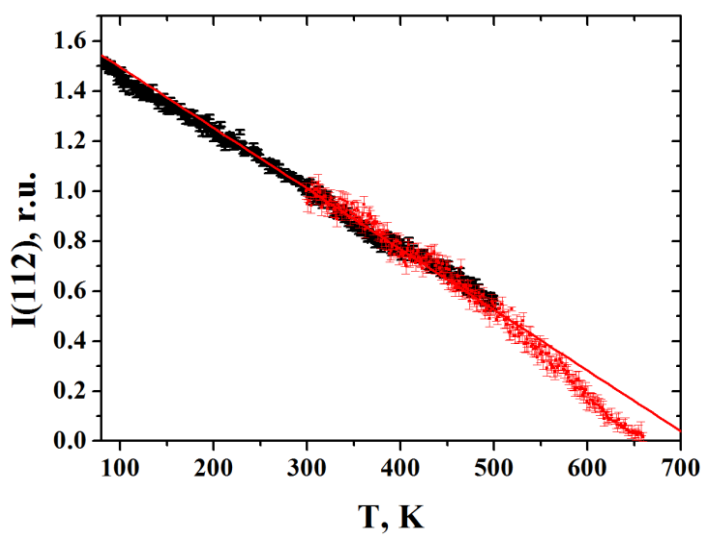

Figure 13. Temperature dependence of the intensity of the 112 peak.

The linear coefficients of thermal expansion defined as $\alpha = \frac{1}{V_0} \cdot \frac{V-V_0}{T-T_0}$, where $V_0$ ($V$) and $T_0$ ($T$) being the initial (final) unit cell volume and the sample temperature, are $5.08 \cdot 10^{-5}$ (T interval 100 - 300 K) for α-$BaFe_2Se_3$, $3.39 \cdot 10^{-5}$ K$^{-1}$ (T interval 400 - 500 K) for β-$BaFe_2Se_3$ and $5.58 \cdot 10^{-5}$ K$^{-1}$ (T interval 660 - 960 K) for γ-$BaFe_2Se_3$.

**Discussion**

Diffraction experiments presented here revealed that, depending on temperature and pressure conditions, $BaFe_2Se_3$ can be found in α, β or γ forms. The first two polymorphs are isostructural in the sense that they are characterized by the same space group and Wyckoff positions, but differ in terms of structural evolution and in particular with thermal expansion coefficients. The last polymorph found above 60 kbar at room temperature and above 660 K at ambient pressure, is a structure with a more symmetric arrangement of the FeSe double chains and Ba cations. It seems from the diffraction and DSC data that pressure and temperature induced transitions α→γ and β→γ are, within the limit of our experimental resolution, of the second order, while α→β transition is of the first order. Considering the γ phase as parent, one finds that the order parameter (OP) responsible for the distortions present in α- and β-structures relates to $Y_2^+$ Irrep which belongs to the Y-point of the orthorhombic Brillouin zone[22]. Microscopically the OP is expressed as rotation of the FeSe double chains together with a shift of the Ba ions normal to the rotation axis. The α and β forms being both of the same symmetry differ by the value of rotational and shifting components.

Such a combination of second order phase transition with strongly first order isostructural transformation is rather unique and, up to our knowledge, has not been met in an experiment before. There are however theoretical indications that such an unusual structural evolution can be parameterized in frame of the phenomenological theory of phase transitions assuming *high degree of non-linearity* for free energy as a function of OP.[13]

As we just mentioned, the structural antiferrodistortive phase transition $Cmcm(Z_p=2)$-$Pnma(Z_p=4)$ is induced by the single-component OP $Y_2^+$. The OP image group, isomorphous to the $C_i$ point group, contains an "inversion operation" changing the sign of OP (A2a group in notations of [22]). The relevant Landau potential is an one-component order-parameter expansion containing, therefore, only even powers of the order parameter η:

$$\Phi(\eta) = a_1\eta^2 + a_2\eta^4 + a_3\eta^6 + a_4\eta^8 + ... + a_m\eta^{2m} \ . \tag{2}$$

The phase diagrams and physical anomalies corresponding to the potentials with maximal OP degrees m = 8 and m = 10 have been earlier studied in detail by Gufan and Larin[23,24], see also a review of the results in [25].

The equation of state for the 8-degree model (2) and corresponding stability conditions are:

$$\frac{d\Phi}{d\eta} = 2\eta \cdot (a_1 + a_2\eta^2 + a_3\eta^4 + a_4\eta^6) = 0 \ ; \tag{3}$$

$$\frac{d^2\Phi}{d\eta^2} = (2\eta)^{-1} \cdot \frac{d\Phi}{d\eta} + 8\eta^2 \cdot (a_2 + 3a_3\eta^2 + 6a_4\eta^4) \geq 0 \ . \tag{4}$$

It should be underlined that the 8-degree expansion (2) is the lowest degree which allows describing two isostructural low-symmetry phases. At $a_3 \geq 0$, $a_4 > 0$ the equation (3) has only two real solutions: $\eta = 0$ (I) describes a parent high-symmetry phase, and $\eta^2 \neq 0$ (II) corresponds to a distorted low-symmetry structure. However, at $a_3 < 0$ and $a_4 > 0$ (the latter should be positive in any 8-degree model as the highest-degree term, in order to ensure its global stability) two "low-symmetry" real solutions do exist: $II_1$ and $II_2$ differ by the magnitude of the order parameter $\eta \neq 0$.

Stability limits for the parent phase I and two distinct but isostructural phases $II_1$ and $II_2$ one finds from the equation (4); phase transition lines are defined by the energy equality conditions: $\Phi_I = \Phi_{II1,2}$, $\Phi_{II1} = \Phi_{II2}$. Figure 14 shows the phase diagram in the ($a_1$-$a_2$) plane, corresponding to the model (2), and the order parameter $\eta$ as a function of the phenomenological variable $a_1$ for the representative thermodynamic paths.

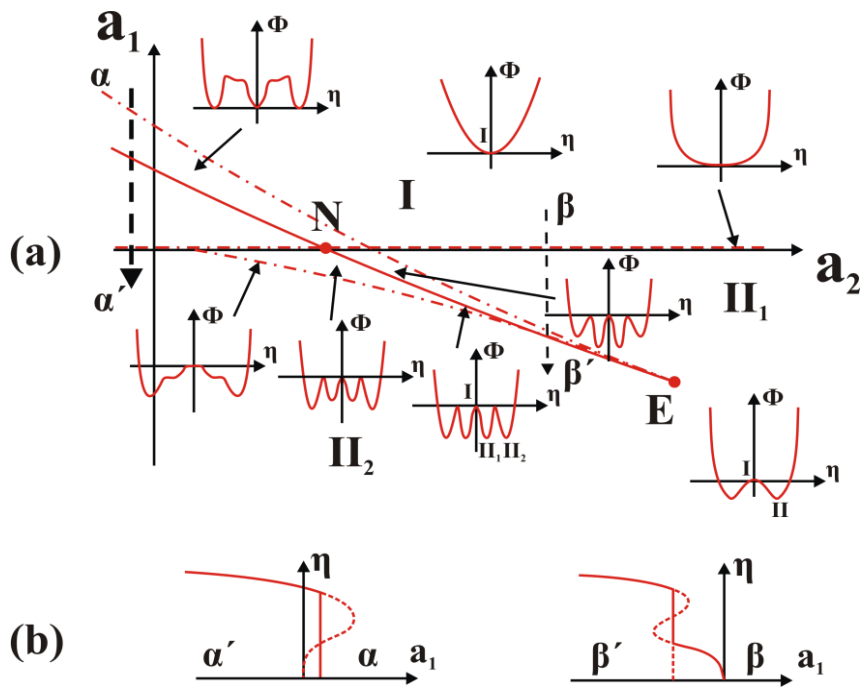

Figure 14. (a) Phase diagram corresponding to the 8-degree potential (2) with $a_3 < 0$ and $a_4 > 0$. Solid, dashed and dashed-dotted lines represent first-order, second-order transitions and stability limit lines, respectively. The Landau potential $\Phi(\eta)$ curves are shown in some significant regions of the phase diagram. N is a triple point, E is a critical end point. (b) Order-parameter variation on different thermodynamic paths: $\alpha\alpha'$- first-order I-$II_2$ transition; $\beta\beta'$- second-order I-$II_1$ transition followed by the first-order isostructural $II_1$-$II_2$ phase transition. The scheme is adopted from Ref. [25].

The second-order transition line between I and II$_1$ ($\Phi_I = \Phi_{III}$) $a_1 = 0$ starts at a triple point N{ $a_1 = 0$, $a_2 = a_3^2/4a_4$}. The line of the first-order transitions I-II$_2$ is:

$$a_1^{eq} = \frac{a_3}{a_4} \cdot a_2^{eq} + \left[ -\frac{4}{27} \cdot \frac{(a_2^{eq})^3}{a_4} \right]^{1/2},$$

$$a_1^{eq} = a_1 - \frac{a_3}{9a_4^2}, \qquad a_2^{eq} = a_2 - \frac{a_3^2}{3a_4}. \qquad (5)$$

The isostructural transition line II$_1$-II$_2$ is a straight-line segment (NE) of equation:

$$a_1 = \frac{a_3}{2a_4} a_2 - \frac{a_3^3}{9a_4^2}. \qquad (6)$$

It is tangent to the first-order transition line I-II$_2$ (5), and ends at the critical point E{$a_1^E = -a_3^3/16a_4^2$, $a_2^E = 3a_3^2/8a_4$}. The OP jump at the isostructural phase transition (on the line NE) corresponds to:

$$\Delta(\eta^2) = \frac{1}{3a_4} \cdot \left[ -a_3 + \left( a_3^2 - 3a_2 a_4 \right)^{1/2} \right]. \qquad (7)$$

In order to map the special (phase transition) points and lines experimentally measured in the P-T coordinate system onto the 2D space of the phenomenological parameters $a_1$-$a_2$, it is convenient to call a linear transformation for

$$a_1(T,P) = \alpha_1(T - T_0) + \beta_1(P - P_0),$$
$$a_2(T,P) = \alpha_2(T - T_0) + \beta_2(P - P_0). \qquad (8)$$

This establishes the (P,T)↔($a_1$,$a_2$) correspondence, using a minimal number of parameters: $\alpha_i$, $\beta_i$, $T_0$, and $P_0$. Both coordinate systems are shown in Fig. 15 along with transformation points measured in our experiments.

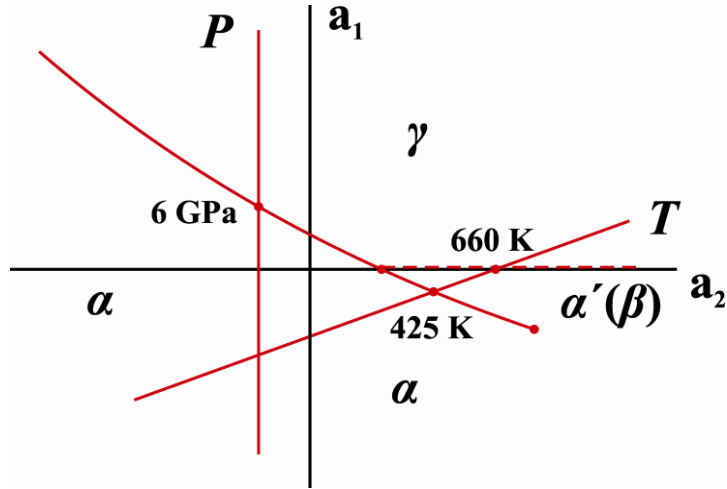

Figure 15. Position of phase transition points on the schematic P-T and $a_1$-$a_2$ phase diagrams of $BaFe_2Se_3$.

One concludes that the singular points observed in our experiments fit well to the topology of the phase diagram which resulted from minimization of the phenomenological Landau potential (2).

As we have shown experimentally, microscopic realization of the order parameter appears as rotations of the $FeSe_4$ tetrahedra combined with displacements of the Ba cations. There is no symmetry restriction for both deformations to respond similar on an external stimulus; they both correspond to the same Irrep as well as any mixture of them. However, it is intuitively clear that a rotation of infinite double chains and shifts of cations filling the space between the chains would have different effects in terms of the internal energy and also in terms of the crystal volume affected by such distortions. As we have shown above, in order to parameterize such a complex response one has to keep high order terms in the series expansion of the free energy; thus appearance of both rotational and shifting components of the OP may require a highly non-linear interaction potential. Still, a nature of this nonlinearity in terms of local interactions has to be uncovered by *ab initio* calculations.

**Conclusions**

We report two new modifications of $BaFe_2Se_3$ that can be found at different thermodynamic conditions. We characterized all of them with the help of single crystal and powder diffraction techniques. A symmetry based analysis of the relationship between different forms is augmented by detailed structural information on the pathways connecting the different phases. The phase transitions are described with the help of phenomenological theory and a generic phase diagram mapping. To the best of our knowledge, this is the first reliable observation of coupled structural and isostructural phase transitions induced by the same symmetry-breaking order parameter.

We show that at least thermal expansion and compressibility differ for different forms of BaFe$_2$Se$_3$. In the case of a potential stabilisation of the β and γ forms by doping, one may expect that other physical properties including diamagnetic response at 11 K would also be affected by polymorphism; however, a range of existence of different phases as a function of chemical doping is still an open question.

**Acknowledgements**

Volodymyr Svitlyk acknowledges support by the Alexander von Humboldt Foundation and the Federal Ministry for Education and Research.

**Supplementary information**

Table 1. Mode amplitudes (m.a.*) and the corresponding displacements maxima ($d_{max}$, Å) for the BaFe$_2$Se$_3$ *Pnma-Cmcm* structural transition from diffraction data at 150 K (single crystal) and 70 kbar (powder).

| Mode | m.a. | $d_{max}$ |
|---|---|---|
| GM1+[BA1:c]A1(a) | 0.247(2) | 0.1234(10) |
| GM1+[FE1:e]A(a) | 0.086(4) | 0.0304(14) |
| GM1+[SE1:c]A1(a) | -0.184(3) | 0.0920(15) |
| GM1+[SE2:g]A'_1(a) | 0.114(3) | 0.0403(11) |
| GM1+[SE2:g]A'_2(a) | 0.008(3) | 0.0028(11) |
| Y2+[BA1:c]B2(a) | -0.3907(19) | 0.1953(9) |
| Y2+[FE1:e]B_1(a) | 0.193(5) | 0.0682(18) |
| Y2+[FE1:e]B_2(a) | 0.000(4) | 0 |
| Y2+[SE1:c]B2(a) | 0.172(3) | 0.0860(15) |
| Y2+[SE2:g]A'_1(a) | 0.578(3) | 0.2044(11) |
| Y2+[SE2:g]A'_2(a) | -0.675(3) | 0.2386(11) |

* The magnitude of the m.a. is the square root of the sum of the squares of the mode-induced changes within the primitive supercell

Table 2. Displacive mode definitions for the BaFe$_2$Se$_3$ *Pnma-Cmcm* structural transition from diffraction data at 150 K (single crystal) and 70 kbar (powder).

| Mode amplitudes/ Atoms | x | y | z | dx | dy | dz |
|---|---|---|---|---|---|---|
| *Cmcm*[0,0,0]GM1+(a)[BA1:c]A1(a) | | | | | | |
| Ba1 | 0.6741 | 0.25 | 0 | 1 | 0 | 0 |
| | 0.1741 | 0.25 | 0.5 | 1 | 0 | 0 |
| | 0.3259 | 0.75 | 0 | -1 | 0 | 0 |
| | 0.8259 | 0.75 | 0.5 | -1 | 0 | 0 |
| *Cmcm*[0,0,0]GM1+(a)[FE1:e]A(a) | | | | | | |
| Fe5 | 0 | 0 | 0.8491 | 0 | 0 | 1 |
| | 0.5 | 0 | 0.3491 | 0 | 0 | 1 |
| | 0 | 0.5 | 0.1509 | 0 | 0 | -1 |
| | 0.5 | 0.5 | 0.6509 | 0 | 0 | -1 |
| | 0 | 0 | 0.1509 | 0 | 0 | -1 |
| | 0.5 | 0 | 0.6509 | 0 | 0 | -1 |
| | 0 | 0.5 | 0.8491 | 0 | 0 | 1 |
| | 0.5 | 0.5 | 0.3491 | 0 | 0 | 1 |
| *Cmcm*[0,0,0]GM1+(a)[SE1:c]A1(a) | | | | | | |
| Se3 | 0.1369 | 0.25 | 0 | 1 | 0 | 0 |
| | 0.6369 | 0.25 | 0.5 | 1 | 0 | 0 |
| | 0.8631 | 0.75 | 0 | -1 | 0 | 0 |
| | 0.3631 | 0.75 | 0.5 | -1 | 0 | 0 |

*Cmcm*[0,0,0]GM1+(a)[SE2:g]A'_1(a)

| | | | | | | |
|---|---|---|---|---|---|---|
| Se2 | 0.3785 | 0.25 | 0.2029 | 0 | 0 | 1 |
| | 0.1215 | 0.75 | 0.7029 | 0 | 0 | 1 |
| | 0.8785 | 0.25 | 0.2971 | 0 | 0 | -1 |
| | 0.6215 | 0.75 | 0.7971 | 0 | 0 | -1 |
| Se4 | 0.8785 | 0.25 | 0.7029 | 0 | 0 | 1 |
| | 0.6215 | 0.75 | 0.2029 | 0 | 0 | 1 |
| | 0.3785 | 0.25 | 0.7971 | 0 | 0 | -1 |
| | 0.1215 | 0.75 | 0.2971 | 0 | 0 | -1 |

*Cmcm*[0,0,0]GM1+(a)[SE2:g]A'_2(a)

| | | | | | | |
|---|---|---|---|---|---|---|
| Se2 | 0.3785 | 0.25 | 0.2029 | 1 | 0 | 0 |
| | 0.1215 | 0.75 | 0.7029 | -1 | 0 | 0 |
| | 0.8785 | 0.25 | 0.2971 | 1 | 0 | 0 |
| | 0.6215 | 0.75 | 0.7971 | -1 | 0 | 0 |
| Se4 | 0.8785 | 0.25 | 0.7029 | 1 | 0 | 0 |
| | 0.6215 | 0.75 | 0.2029 | -1 | 0 | 0 |
| | 0.3785 | 0.25 | 0.7971 | 1 | 0 | 0 |
| | 0.1215 | 0.75 | 0.2971 | -1 | 0 | 0 |

*Cmcm*[1,0,0]Y2+(a)[BA1:c]B2(a)

| | | | | | | |
|---|---|---|---|---|---|---|
| Ba1 | 0.6741 | 0.25 | 0 | 0 | 0 | 1 |
| | 0.1741 | 0.25 | 0.5 | 0 | 0 | -1 |
| | 0.3259 | 0.75 | 0 | 0 | 0 | -1 |
| | 0.8259 | 0.75 | 0.5 | 0 | 0 | 1 |

*Cmcm*[1,0,0]Y2+(a)[FE1:e]B_1(a)

| | | | | | | |
|---|---|---|---|---|---|---|
| Fe5 | 0 | 0 | 0.8491 | 1 | 0 | 0 |
| | 0.5 | 0 | 0.3491 | -1 | 0 | 0 |
| | 0 | 0.5 | 0.1509 | -1 | 0 | 0 |
| | 0.5 | 0.5 | 0.6509 | 1 | 0 | 0 |
| | 0 | 0 | 0.1509 | -1 | 0 | 0 |
| | 0.5 | 0 | 0.6509 | 1 | 0 | 0 |
| | 0 | 0.5 | 0.8491 | 1 | 0 | 0 |
| | 0.5 | 0.5 | 0.3491 | -1 | 0 | 0 |

*Cmcm*[1,0,0]Y2+(a)[FE1:e]B_2(a)

| | | | | | | |
|---|---|---|---|---|---|---|
| Fe5 | 0 | 0 | 0.8491 | 0 | 1 | 0 |
| | 0.5 | 0 | 0.3491 | 0 | -1 | 0 |
| | 0 | 0.5 | 0.1509 | 0 | 1 | 0 |
| | 0.5 | 0.5 | 0.6509 | 0 | -1 | 0 |
|  | 0 | 0 | 0.1509 | 0 | -1 | 0 |
|  | 0.5 | 0 | 0.6509 | 0 | 1 | 0 |
|  | 0 | 0.5 | 0.8491 | 0 | -1 | 0 |
|  | 0.5 | 0.5 | 0.3491 | 0 | 1 | 0 |

*Cmcm*[1,0,0]Y2+(a)[SE1:c]B2(a)

| Se3 | 0.1369 | 0.25 | 0 | 0 | 0 | 1 |
|---|---|---|---|---|---|---|
|  | 0.6369 | 0.25 | 0.5 | 0 | 0 | -1 |
|  | 0.8631 | 0.75 | 0 | 0 | 0 | -1 |
|  | 0.3631 | 0.75 | 0.5 | 0 | 0 | 1 |

*Cmcm*[1,0,0]Y2+(a)[SE2:g]A'_1(a)

| Se2 | 0.3785 | 0.25 | 0.2029 | 0 | 0 | 1 |
|---|---|---|---|---|---|---|
|  | 0.1215 | 0.75 | 0.7029 | 0 | 0 | 1 |
|  | 0.8785 | 0.25 | 0.2971 | 0 | 0 | -1 |
|  | 0.6215 | 0.75 | 0.7971 | 0 | 0 | -1 |
| Se4 | 0.8785 | 0.25 | 0.7029 | 0 | 0 | -1 |
|  | 0.6215 | 0.75 | 0.2029 | 0 | 0 | -1 |
|  | 0.3785 | 0.25 | 0.7971 | 0 | 0 | 1 |
|  | 0.1215 | 0.75 | 0.2971 | 0 | 0 | 1 |

*Cmcm*[1,0,0]Y2+(a)[SE2:g]A'_2(a)

| Se2 | 0.3785 | 0.25 | 0.2029 | 1 | 0 | 0 |
|---|---|---|---|---|---|---|
|  | 0.1215 | 0.75 | 0.7029 | -1 | 0 | 0 |
|  | 0.8785 | 0.25 | 0.2971 | 1 | 0 | 0 |
|  | 0.6215 | 0.75 | 0.7971 | -1 | 0 | 0 |
| Se4 | 0.8785 | 0.25 | 0.7029 | -1 | 0 | 0 |
|  | 0.6215 | 0.75 | 0.2029 | 1 | 0 | 0 |
|  | 0.3785 | 0.25 | 0.7971 | -1 | 0 | 0 |
|  | 0.1215 | 0.75 | 0.2971 | 1 | 0 | 0 |

Table 3. Mode amplitudes (m.a.*) and the corresponding displacements maxima ($d_{max}$, Å) for temperature induced structural changes of $BaFe_2Se_3$ from single crystal diffraction data at 150 and 480 K

| Mode | m.a. | $d_{max}$ |
|---|---|---|
| GM1+[Ba1:c]A'_1(a) | 0.040(2) | 0.0200(10) |
| GM1+[Ba1:c]A'_2(a) | 0.140(2) | 0.0700(10) |
| GM1+[Se2:c]A'_1(a) | -0.142(4) | 0.071(2) |
| GM1+[Se2:c]A'_2(a) | 0.120(3) | 0.0600(15) |
| GM1+[Se3:c]A'_1(a) | 0.065(4) | 0.033(2) |
| GM1+[Se3:c]A'_2(a) | -0.051(3) | 0.0255(15) |
| GM1+[Se4:c]A'_1(a) | 0.148(4) | 0.074(2) |
| GM1+[Se4:c]A'_2(a) | 0.110(4) | 0.0550(20) |

| | | | | | | |
|---|---|---|---|---|---|---|
| GM1+[Fe5:d]A_1(a) | | | | -0.059(5) | | 0.0209(18) |
| GM1+[Fe5:d]A_2(a) | | | | 0.000(5) | | 0 |
| GM1+[Fe5:d]A_3(a) | | | | 0.053(5) | | 0.0187(18) |

*The magnitude of the m.a. is the square root of the sum of the squares of the mode-induced changes within the primitive supercell

Table 4. Displacive mode definitions for temperature induced structural changes of BaFe$_2$Se$_3$ from single crystal diffraction data at 150 and 480 K

| Mode amplitudes/ Atoms | x | y | z | dx | dy | dz |
|---|---|---|---|---|---|---|
| *Pnma*[0,0,0]GM1+(a)[Ba1:c]A'_1(a) | | | | | | |
| Ba1 | 0.18378 | 0.25 | 0.51469 | 1 | 0 | 0 |
| | 0.68378 | 0.25 | 0.98531 | 1 | 0 | 0 |
| | 0.81622 | 0.75 | 0.48531 | -1 | 0 | 0 |
| | 0.31622 | 0.75 | 0.01469 | -1 | 0 | 0 |
| | | | | | | |
| *Pnma*[0,0,0]GM1+(a)[Ba1:c]A'_2(a) | | | | | | |
| Ba1 | 0.18378 | 0.25 | 0.51469 | 0 | 0 | 1 |
| | 0.68378 | 0.25 | 0.98531 | 0 | 0 | -1 |
| | 0.81622 | 0.75 | 0.48531 | 0 | 0 | -1 |
| | 0.31622 | 0.75 | 0.01469 | 0 | 0 | 1 |
| | | | | | | |
| *Pnma*[0,0,0]GM1+(a)[Se2:c]A'_1(a) | | | | | | |
| Se2 | 0.36277 | 0.25 | 0.2243 | 1 | 0 | 0 |
| | 0.86277 | 0.25 | 0.2757 | 1 | 0 | 0 |
| | 0.63723 | 0.75 | 0.7757 | -1 | 0 | 0 |
| | 0.13723 | 0.75 | 0.7243 | -1 | 0 | 0 |
| | | | | | | |
| *Pnma*[0,0,0]GM1+(a)[Se2:c]A'_2(a) | | | | | | |
| Se2 | 0.36277 | 0.25 | 0.2243 | 0 | 0 | 1 |
| | 0.86277 | 0.25 | 0.2757 | 0 | 0 | -1 |
| | 0.63723 | 0.75 | 0.7757 | 0 | 0 | -1 |
| | 0.13723 | 0.75 | 0.7243 | 0 | 0 | 1 |
| | | | | | | |
| *Pnma*[0,0,0]GM1+(a)[Se3:c]A'_1(a) | | | | | | |
| Se3 | 0.62575 | 0.25 | 0.49299 | 1 | 0 | 0 |
| | 0.12575 | 0.25 | 0.00701 | 1 | 0 | 0 |
| | 0.37425 | 0.75 | 0.50701 | -1 | 0 | 0 |
| | 0.87425 | 0.75 | 0.99299 | -1 | 0 | 0 |
| | | | | | | |
| *Pnma*[0,0,0]GM1+(a)[Se3:c]A'_2(a) | | | | | | |
| Se3 | 0.62575 | 0.25 | 0.49299 | 0 | 0 | 1 |
| | 0.12575 | 0.25 | 0.00701 | 0 | 0 | -1 |
| | 0.37425 | 0.75 | 0.50701 | 0 | 0 | -1 |
|  | 0.87425 | 0.75 | 0.99299 | 0 | 0 | 1 |
| *Pnma*[0,0,0]GM1+(a)[Se4:c]A'_1(a) | | | | | | |
| Se4 | 0.39452 | 0.25 | 0.80986 | 1 | 0 | 0 |
|  | 0.89452 | 0.25 | 0.69014 | 1 | 0 | 0 |
|  | 0.60548 | 0.75 | 0.19014 | -1 | 0 | 0 |
|  | 0.10548 | 0.75 | 0.30986 | -1 | 0 | 0 |
| *Pnma*[0,0,0]GM1+(a)[Se4:c]A'_2(a) | | | | | | |
| Se4 | 0.39452 | 0.25 | 0.80986 | 0 | 0 | 1 |
|  | 0.89452 | 0.25 | 0.69014 | 0 | 0 | -1 |
|  | 0.60548 | 0.75 | 0.19014 | 0 | 0 | -1 |
|  | 0.10548 | 0.75 | 0.30986 | 0 | 0 | 1 |
| *Pnma*[0,0,0]GM1+(a)[Fe5:d]A_1(a) | | | | | | |
| Fe5 | 0.49549 | 0 | 0.35056 | 1 | 0 | 0 |
|  | 0.99549 | 0.5 | 0.14944 | 1 | 0 | 0 |
|  | 0.50451 | 0.5 | 0.64944 | -1 | 0 | 0 |
|  | 0.00451 | 0 | 0.85056 | -1 | 0 | 0 |
|  | 0.50451 | 0 | 0.64944 | -1 | 0 | 0 |
|  | 0.00451 | 0.5 | 0.85056 | -1 | 0 | 0 |
|  | 0.49549 | 0.5 | 0.35056 | 1 | 0 | 0 |
|  | 0.99549 | 0 | 0.14944 | 1 | 0 | 0 |
| *Pnma*[0,0,0]GM1+(a)[Fe5:d]A_2(a) | | | | | | |
| Fe5 | 0.49549 | 0 | 0.35056 | 0 | 1 | 0 |
|  | 0.99549 | 0.5 | 0.14944 | 0 | -1 | 0 |
|  | 0.50451 | 0.5 | 0.64944 | 0 | 1 | 0 |
|  | 0.00451 | 0 | 0.85056 | 0 | -1 | 0 |
|  | 0.50451 | 0 | 0.64944 | 0 | -1 | 0 |
|  | 0.00451 | 0.5 | 0.85056 | 0 | 1 | 0 |
|  | 0.49549 | 0.5 | 0.35056 | 0 | -1 | 0 |
|  | 0.99549 | 0 | 0.14944 | 0 | 1 | 0 |
| *Pnma*[0,0,0]GM1+(a)[Fe5:d]A_3(a) | | | | | | |
| Fe5 | 0.49549 | 0 | 0.35056 | 0 | 0 | 1 |
|  | 0.99549 | 0.5 | 0.14944 | 0 | 0 | -1 |
|  | 0.50451 | 0.5 | 0.64944 | 0 | 0 | -1 |
|  | 0.00451 | 0 | 0.85056 | 0 | 0 | 1 |
|  | 0.50451 | 0 | 0.64944 | 0 | 0 | -1 |
|  | 0.00451 | 0.5 | 0.85056 | 0 | 0 | 1 |
|  | 0.49549 | 0.5 | 0.35056 | 0 | 0 | 1 |
|  | 0.99549 | 0 | 0.14944 | 0 | 0 | -1 |